\documentstyle [12pt] {article}

\newcommand{\be}{ \begin{eqnarray}}
\newcommand{\ee}{\end{eqnarray}}

\begin{document}

\title{ Event by Event fluctuations and Inclusive Distributions}
\author{ A. Bialas\thanks{e-mail: bialas@thrisc.if.uj.edu.pl}\\
M.Smoluchowski Institute of Physics \\ Jagellonian University, Cracow \\
\and V. Koch\thanks{e-mail: vkoch@lbl.gov}\\
 Lawrence Berkeley National Laboratory, \\
Berkeley, CA 94720, U.S.A.}
\maketitle
\vspace{-10cm}
\hfill LBNL-42848,  TPJU-2/99
\vspace{10cm}
\abstract {Event-by-event observables are compared with conventional inclusive
  measurements. We find that moments of  event-by-event fluctuations 
  are closely
  related to inclusive correlation functions. Implications for upcomming heavy
  ion experiments are discussed.}

\ \\
\ \\
{\bf 1.} It is now widely recognized that studies of  event-by-event
fluctuations
observed in  high energy multi-particle reactions may
become an important tool in
 attempts to understand the underlying dynamics of ultrarelativistic heavy ion
 collisions \cite{gm} - \cite{Shu98}. It has for instance been proposed
that the
 measurement of event-by-event fluctuations of the temperature via e.g. the
 transverse momentum spectrum could provide information about the heat
 capacity of the system generated in these collisions \cite{Sto95,Shu98}.
Also, by investigating event-by-event fluctuations one may be able to
determine the degree of thermalization of the system \cite{gm,g} or to
select distinct event classes.
The analysis of heavy ion collisions on an event-by-event basis has been
pioneered by the NA49 experiment. First, preliminary results \cite{NA49} seem
to indicate that the observed fluctuation in the mean transverse momentum as
well as the kaon to pion ratio are of Gaussian shape.

It seems, therefore, interesting
to study in some detail what is the information content of such measurements
and to what extent they actually
 differ from
the more conventional treatment of particle  spectra.
In the present paper we discuss the relation of event-by-event fluctuations
to the
standard inclusive (multi-particle) distributions. Our  conclusions can be
summarized as follows: Moments of event-by-event fluctuations of any
(multi-particle) observable can be expressed in terms of inclusive
distributions, provided the inclusive  distributions are known up to twice the
order of the observable under consideration. 
For instance, in order to express the dispersion of the event-by-event
distribution
of a single
particle
observable, one needs to know the two-particle inclusive distribution
etc.
Fluctuations of ratios of observables, on the other hand,
cannot simply be expressed in terms of ratios of inclusive measurements.
In particular, to obtain moments of an observable corresponding to an
"intensive" quantity (in the thermodynamic limit) it is necessary to know the inclusive distributions in narrow multiplicity intervals.
However, as long as the central limit theorem can be applied, i.e. the
observables involved are dominated by independent single particle emission,
and the observed multiplicities are reasonably high, the knowledge of
inclusive distribution of twice the order of the observables under
consideration again is sufficient.

\vspace{0.5cm}

{\bf 2.} Let us consider a variable $x(p)$ which depends on momentum of
one particle. We shall discuss
event-by-event fluctuations of the quantity
\begin{equation}
S(x) = \sum_{i=1}^N x(p_i) \equiv \sum_{i=1}^N x(i)   \label{3a}
\end{equation}
where N is the multiplicity of the  event.

The {\em event averaged}
moments of this quantity can be expressed as
\begin{equation}
<S^k>= \frac1{M} \sum_{m=1}^{M}
\sum_{i_1=1}^{N_m}...\sum_{i_k=1}^{N_m} x_m(i_1)...x_m(i_k)
\label{4a}
\end{equation}
where $m$ labels the different events and
$M$ is their  total number. $N_m$ is the multiplicity of the event labelled by
$m$.

On the other hand, the moments of $x(p)$ calculated from n-particle inclusive
distribution $\rho_n(p_1,...,p_n)$
are defined as
\begin{eqnarray}
\int dp_1...dp_n \rho_n(p_1,...,p_n) [x(p_1)]^{k_1}...
[x(p_n)]^{k_n}  = \nonumber \\  \frac1{M}
\sum_{m=1}^M\sum_{i_1=1}^{N_m}...\sum_{i_n=1}^{N_m}
[x_m(i_1)]^{k_1}...[x_m(i_n)]^{k_n}
\label{2a}
\end{eqnarray}
where the sums over $i_1...i_n$  include only the terms for which all indices
$i_1...i_n$ are
different from each other.

One sees immediately that (\ref{4a}) and (\ref{2a}) are related.
\begin{equation}
<S> = \int dp \rho_1(p) x(p)   \label{3}
\end{equation}
\begin{equation}
<S^2> = \int dp_1 dp_2 \rho_2(p_1,p_2) \, x(p_1) x(p_2) 
+ \int d p \rho_1(p) \, [x(p)]^2   \label{4}
\end{equation}
\begin{eqnarray}
<S^3> & = & \int dp_1 dp_2 dp_3 \rho_3(p_1,p_2,p_3) \,x(p_1)x(p_2)x(p_3)
\nonumber \\ 
& & \mbox{}+ 3 \int dp_1dp_2 \rho_2(p_1,p_2) \,x(p_1) [x(p_2)]^2 
\nonumber \\
& & \mbox{} + \int dp \rho_1(p) \, [x(p)]^3    \label{5}
\end{eqnarray}
Similar formulae can be derived for higher moments of $S$.

We have thus established the relation between inclusive measurements and
event-by-event fluctuations
for single particle observables \cite{gm}, such as e.g. the
transverse momentum, particle abundances \cite{g} etc.

\vspace{.5cm}

{\bf 3.} The same argument can be constructed for variables which depend on
two or more
particle momenta. In particular the fluctuations of Hanburry-Brown Twiss
(HBT) two particle correlations belong to this class. They are of practical
interest and will be investigated in future heavy ion experiments \cite{Star}.
Here we will restrict the argument to two particles but it can be readily
extended to multiparticle correlations.
Consider a variable
 $ y = y(p,p')$. We calculate the event by event
fluctuation of the quantity
\begin{equation}
T \equiv \sum_{i=1}^N\sum_{j=1}^N y (p_i,p_j)   \label{6}
\end{equation}
where $N$ is the multiplicity of the event and the sum runs only for $i\neq j$
($y$ is not
defined for $i=j$).

Similarly as before we thus can write
\begin{equation}
<T^k>=\frac1{M} \sum_{m=1}^M
\sum_{i_1=1}^{N_m}\sum_{j_1=1}^{N_m}...\sum_{i_k=1}^{N_m}\sum_{j_k=1}^{N_m}
y_m(i_1,j_1)...y_m(i_k,j_k)     \label{7}
\end{equation}
Consequently we obtain
\begin{equation}
<T> = \int dp_1 dp_2 \rho_2(p_1,p_2) y(p_1,p_2)   \label{8}
\end{equation}
and
\begin{eqnarray}
<T^2> & = &   
\int dp_1dp_2dp_3dp_4 \rho_4(p_1,p_2,p_3,p_4) \, y(p_1,p_2) y(p_3,p_4) 
\nonumber \\ 
& & \mbox{}+  4 \int dp_1 dp_2 dp_3 \rho_3(p_1,p_2,p_3) \,y(p_1,p_2) y(p_1,p_3)
\nonumber \\
& & \mbox{} + 2\int dp_1 dp_2 \rho_2(p_1,p_2) \, [y(p_1,p_2)]^2   \label{9}
\end{eqnarray}
and similarly for higher moments.
Thus for any (multiparticle)
observable
event-by-event fluctuations can be re-expressed in terms of inclusive
multiparticle distribution. The multiparticle distributions need to be known up
to twice the order of the observable under consideration.

\vspace{0.5cm}

{\bf 4.} The argument of Sections 2 and 3  shows that the inclusive measurements
give a precise information on event-by-event fluctuations of the quantities $S$
and $T$
defined by (\ref{3a}) and (\ref{6}). As shown in \cite{gm,g,m} they may be
very useful in investigation of the properties of the multipaticle system.
It is also often  interesting, however,  to
discuss the averages, as they resemble  intensive variables in the
thermodynamic limit:
\begin{equation}
s\equiv S/N;\;\;\;\;  t\equiv \frac{T}{N(N-1)}  \label{10}
\end{equation}
Unfortunately, they can only be obtained if one knows the  inclusive
distribution separately
for each multiplicity or in other words if one knows the inclusive
distributions of the entire ratios and their second moments.
The relevant formulae are easily obtained from those in section 2 and 3.

An approximate estimate can, however,
be obtained from sole inclusive spectra, provided the correlations between the
produced particles are not too strong. This can be seen as follows.

If the correlations are not overwhelmingly strong,  the
dispersion of $S$ at fixed $N$ is likely to follow  the central limit
theorem, i.e.
\begin{equation}
D^2_N \equiv <S^2>_N -<S>_N^2 \simeq N \delta^2            \label{1b}
\end{equation}
where $\delta$ is a constant, independent of $N$. Assume furthermore that
the average value of $x$
does not
depend on multiplicity. Then we have
\begin{equation}
<S>_N \equiv  N<s>_N = N \sigma     \label{2b}
\end{equation}
where $\sigma$ does not depend on multiplicity. Consequently,
\begin{equation}
<S^2>_N= N^2 \sigma^2 + N \delta^2    \label{3b}
\end{equation}
(\ref{2b}) and (\ref{3b}) can be now rewritten in terms of inclusive quantities:

\begin{equation}
<S> \equiv \sum P_N <S>_N \simeq <N> \sigma   \label{4b}
\end{equation}
\begin{equation}
<S^2> \equiv \sum P_N <S^2>_N = <N^2> \sigma^2 + <N> \delta^2  \label{5b}
\end{equation}
Using \ref{3} and \ref{4} we can thus find $\sigma$ and $\delta^2$ from the
inclusive measurements, provided we know
$<N^2>$ and $<N>$. But they are actually known, because they can be obtained
from inclusive spectra
\begin{equation}
<N>=\int \rho(p) dp;\;\;\;\; <N(N-1)> = \int dp_1 dp_2 \rho(p_1,p_2)
\label{6b}
\end{equation}
The question now is can we express $<s>$ and $<s^2>$ in terms of $\sigma$ and
$\delta^2$. The answer is: almost.

We have:
\begin{equation}
<s>_N= \sigma \;\;\;\; \rightarrow <s> \equiv \sum P_N <s>_N =  \sigma
\label{7b}
\end{equation}
This is easy. But from (\ref{3b}) we have
\begin{equation}
<s^2>_N \equiv <S^2>_N/N^2 = \sigma^2 + \delta^2/N   \label{8b}
\end{equation}
and thus
\begin{equation}
<s^2>-<s>^2 = \delta^2 <\frac1{N}>    \label{9b}
\end{equation}
and one sees that we need the average of $1/N$ which is not possible to
obtain if only two-arm spectrometer is available. But  this is not a
serious problem in practice.

Indeed, as long as the fluctuations in N are small
$<\frac{1}{N}>$ to a good approximation is given by
\begin{equation}
<\frac{1}{N}> = \frac{1}{M} \sum_{i=1}^{M} \frac{1}{N_i} \simeq
\frac{1}{<N>} ( 1 + \frac{<(N - <N>)^2>}{<N>^2})
\end{equation}
For typical `central' trigger conditions at a 200 AGeV $Pb+Pb$ collisions,
there are about 200 negatively charged particles per unit rapidity at
mid rapidity. Assuming that a reasonable two arm spectrometer covers about
half a unit of rapidity the
fluctuations in the number of particles are, assuming simple statistics, about
\begin{equation}
\frac{<(N - <N>)^2>}{<N>^2} \simeq \frac{1}{<N>} \simeq 1 \%
\end{equation}

\vspace{.05cm}

{\bf 5.}  We would like to close this paper with the following comments.

(i)  Our argument shows that there is a close link between the
event-by-event analysis and the
inclusive measurements. It follows that, contrary to common belief,
accuracy of the event-by-event analysis hardly depends on
event multiplicity and thus can be useful even for low multiplicity
events. Of course the approximation for $<1/N>$ presented above gets less
reliable with small multiplicities. Also the applicability of the central
limit theorem is less reliable for small multiplicities.
However, for fixed multiplicity, our arguments are rigorous and thus the
differences between and event-by-event analysis and the inclusive measurement
can be minimized by narrow multiplicity cuts.
Furthermore, one should remember that these differences are solely due to the
fact that one wants to express the ratio of observables in terms of ratios of
inclusive expectation values. In principle this is not necessary and the ratio
itself can be obtained via an inclusive measurement. In that case our
arguments again are rigorous.

(ii) It is important to realize that the multiplicity $N$ which enters the
formulae of this paper
is {\it not necessarily} the total multiplicity of the event.
It is the multiplicity of the
particles {\it of interest} and refers to the specific phase-space region under
investigation. In that case, of course, also the inclusive densities
refer to the same particles.
This means in particular that it is not necessary to use a  $4\pi$ detector
in order to apply
the argument of the present paper.

(iii) In practice, the formulae of sections 2 and 3 can only be useful for
the low rank moments.
Nevertheless such information is very often of great interest \cite{gm,g,m}.
Needless
to say, they cannot be  used
to investigate the tail of the event-by-event distribution, i.e. for search
of rare, exotic
events.

(iv) In principle any large acceptance spectrometer, such as NA49, can also be
used as two arm spectrometer. Thus, the above fluctuations can be extracted on
an event-by-event basis as well as in the inclusive way and it would be
interesting to compare the results of both approaches.

In conclusions, we have written down the explicit relation between the
event-by-event analysis and
inclusive multi-particle measurements. The relation is rather straightforward
and can be useful
when applied in experiments where no direct event-by-event measure is possible.

\vspace{.5cm}

{\bf Acknowledgments}

The authors thank the organizers of the Trento workshop on `Event-by-Event
analysis' for the kind hospitality. It was there where this work has been
started.
We would also like to thank S.Mrowczynski, G. Rai, G. Roland and S. Voloshin
for useful discussions and correspondence.
A.B. thanks Service de Physique Theorique de Saclay for a kind
hospitality where part of this work was done. A.B.  was supported in part by the
KBN Grant 2P03B 081614.
VK was supported by the Director, Office of Energy Research,
division of Nuclear Physics of the Office of High Energy and Nuclear
Physics of the U.S. Department of energy under Contract No.
DE-AC03-76SF00098.

\end{document}